\documentclass[aps,prb,longbibliography,twocolumn]{revtex4-2}

\usepackage{graphicx}
\usepackage{dcolumn}
\usepackage{textcomp}
\usepackage[T1]{fontenc}
\usepackage{times}
\usepackage[colorlinks = true, linkcolor = blue, urlcolor  = blue, citecolor = blue, anchorcolor = blue]{hyperref}
\usepackage[english]{babel}
\usepackage{multirow}
\usepackage{dcolumn}

\usepackage{titlesec}

\titlespacing\section{0pt}{18pt plus 4pt minus 2pt}{6pt plus 2pt minus 2pt}
\titlespacing\subsection{0pt}{12pt plus 4pt minus 2pt}{6pt plus 2pt minus 2pt}
\titlespacing\subsubsection{0pt}{12pt plus 4pt minus 2pt}{6pt plus 2pt minus 2pt}

\begin{document}
\title{Quasiparticle band structures of bulk and few-layer PdSe$_2$ from first-principles {\em GW} calculations}
\author{Han-gyu Kim}
\author{Hyoung Joon Choi}
\email{h.j.choi@yonsei.ac.kr}
\affiliation{Department of Physics, Yonsei University, Seoul 03722, Korea}

\begin{abstract}
We performed first-principles density functional theory (DFT) and {\em GW} calculations
to investigate electronic structures of bulk and few-layer PdSe$_2$.
We obtained the quasiparticle band structure of bulk PdSe$_2$, and the obtained energy gap agrees excellently with the reported experimental value. 
For monolayer and bilayer PdSe$_2$, we obtained quasiparticle band structures with respect to the vacuum level.
We analyzed DFT and {\em GW} band structures in detail, finding
$k$-space positions of valence band maxima and conduction band minima,
effective masses, the quasiparticle density of states, work functions, ionization potentials, electron affinities, and $k$-space shapes of electron and hole pockets.
These results provide a foundation for development of basic studies and device applications.
\end{abstract}

\maketitle

\section{\label{sec:sec1}INTRODUCTION}

Two-dimensional (2D) materials have attracted much attention due 
to their 
versatile electronic and magnetic properties and fabrication of heterojunctions
by mechanical exfoliation and stacking. 
While graphene is metallic without any band gap, 
transition-metal dichalcogenides (TMDs) are semiconducting and applied to various device applications due to their finite band gaps \cite{Mak2010,Splendiani2010,Britnell2013,Zhang2014,Kim2021,Oh2019,Park2015}.
Recently, PdSe$_2$, a layered TMD material with stacks of a puckered pentagonal lattice, 
was adopted for device fabrication and showed 
electron-dominant transport with high mobility, a large on-off ratio of current \cite{Chow2017},
and air stability \cite{Oyedele2017}.
PdSe$_2$ is also interesting for the possibility of Stoner-type ferromagnetism in hole-doped 2D layers \cite{SHZhang2018}.

Bulk PdSe$_2$ is a semiconductor with a band gap of 0.4~eV from the resistivity measurement \cite{Hulliger1965}.
A previous density functional theory (DFT) calculation using the local density approximation, however, showed that bulk PdSe$_2$ is a semimetal with no band gap \cite{Hamidani2010}, and a DFT calculation using the modified Becke-Johnson potential showed that bulk PdSe$_2$ is a semiconductor with a very small band gap of 0.03~eV \cite{Sun2015}. 
For accurate study of the band gap, the {\em GW} method \cite{Hedin1965,Strinati1980,Strinati1982,Hybertsen1985,Hybertsen1986} is required.
The {\em GW} method directly computes quasiparticle (QP) energies,
and it is very successful in describing the band gap accurately \cite{Schilfgaarde2006}.
For bulk PdSe$_2$, so far, there is no report on band structures using the {\em GW} method.

For monolayer and bilayer PdSe$_2$, previous DFT calculations \cite{Oyedele2017, Kuklin2019,Lebegue2013} showed that their band gaps are about 1.3 and 0.8~eV, respectively. {\em GW} calculations were also preformed previously, reporting that monolayer and bilayer PdSe$_2$ have QP band gaps of 2.55 and 1.89~eV, respectively \cite{Kuklin2019}.
Meanwhile, band gaps of monolayer and bilayer PdSe$_2$ estimated from optical absorption experiments are close to DFT results \cite{Oyedele2017},
so the difference between experimental band gaps and {\em GW} band gaps is expected to be related to excitonic binding energies, substrate effects, and structural defects such as Se vacancies \cite{Kuklin2019, Ugeda2014}.
For device applications, the accurate study of band-edge energies with respect to the vacuum level is required, which is related to work functions, ionization potentials, and electron affinities. These properties are not reported yet, and can be obtained accurately by the {\em GW} method with careful convergence in 2D geometries.

In the present work,
we study electronic structures of bulk and few-layer PdSe$_2$ 
using DFT and {\em GW} calculations.
We obtain quasiparticle 
band structures, band gaps, the projected density of states, work functions, ionization potentials, and electron affinities for the monolayer, bilayer, and bulk.
We also obtain $k$-space positions of valence band maxima (VBMs) and conduction band minima (CBMs), effective masses at VBMs and CBMs, and $k$-space shapes of electron and hole pockets.

This paper is organized as follows. 
In Sec.~\ref{sec2pdse2}, we describe our calculation method including atomic structures of bulk and few-layer PdSe$_2$ and
parameters for DFT and {\em GW} calculations. 
In Sec.~\ref{sec3pdse2}, we present and discuss 
electronic structures of monolayer, bilayer, and bulk PdSe$_2$ obtained from DFT and {\em GW} calculations.
Finally, we summarize our work in Sec.~\ref{sec4pdse2}.

\section{\label{sec2pdse2}METHODOLOGY}

\subsection{Atomic structures}

\begin{figure}
\includegraphics[scale=1]{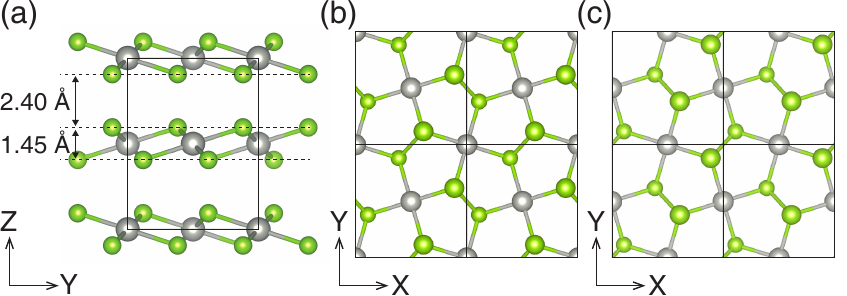}
\caption{\label{fig1pdse2}Atomic structure of bulk PdSe$_2$:
(a) side view, (b) top view of the first triple layer,
and (c) top view of the second triple layer.
Gray and light green dots are Pd and Se atoms, respectively. 
In (a), distances between layers are experimental ones.
The lattice vectors $a, b,$ and $c$ are along the $x, y,$ and $z$ axes, respectively.
In (b) and (c), Se atoms above (below) the Pd plane are represented with larger (smaller) light green dots.}
\end{figure}

We use the experimental atomic structure of 
bulk PdSe$_2$ shown in Fig.~\ref{fig1pdse2}.
Experimental lattice constants are $a$ = 5.7457, $b$ = 5.8678, 
and $c$ = 7.6946~{\AA} \cite{Soulard2004},
and in our work the lattice vectors $a, b,$ and $c$ are along the $x, y,$ and $z$ axes, respectively. 
We construct atomic structures of monolayer and bilayer 
PdSe$_2$ 
by cutting the bulk atomic structure and including
a large enough vacuum region of about 20~{\AA}.
To simulate the vacuum level for bulk PdSe$_2$, we construct 12-layer PdSe$_2$.

The monolayer PdSe$_2$ is a triple layer consisting of three atomic layers of Se, Pd, and Se,
where each Pd atom is neighbored by four Se atoms and each Se atom is neighbored by two Pd atoms and one Se atom, as shown in Figs.~\ref{fig1pdse2}(b) and \ref{fig1pdse2}(c).
The triple layer has a thickness of 1.45 {\AA} and contains two Pd atoms and four Se atoms in its unit cell, as shown in Fig.~\ref{fig1pdse2}. 
The bulk geometry contains two triple layers in the unit cell, as shown in Fig.~\ref{fig1pdse2}(a),
where triple layers are separated by 2.40~{\AA}.
Atomic positions are from experimental values \cite{Soulard2004}; that is, one of the Pd atoms is at (0, 0, 0), and one of the Se atoms is at (0.11125, 0.11799, 0.40573) in terms of fractional coordinates with respect to lattice vectors, and other atomic positions are determined by symmetries.

The bulk geometry belongs to space group 61 ($Pbca$), which has inversion symmetry and three glide-mirror planes consisting of $xy$, $yz$, and $zx$ mirror planes and translation by a half unit cell along the $x, y,$ and $z$ directions, respectively.
The monolayer geometry belongs to space group 14 ($P2{_1}/c$), which has inversion symmetry and one glide-mirror plane consisting of the $yz$ mirror plane and translation by one half of the lattice vector $b$.
The bilayer geometry belongs to space group 7 ($Pc$), which has no inversion symmetry but one glide-mirror plane consisting of the $yz$ mirror plane and translation by one half of the lattice vector $b$.

\subsection{DFT calculations}

We perform DFT calculations \cite{Hohenberg1964, Kohn1965} using the QUANTUM ESPRESSO 
code~\cite{Giannozzi2009} with the Perdew-Burke-Ernzerhof-type 
generalized gradient approximation~\cite{Perdew1996} to 
the exchange-correlation energy. We use a kinetic-energy cutoff of 100 Ry 
for the plane wave and norm-conserving pseudopotentials. 
The norm-conserving pseudopotential for Pd is constructed by including 
semicore $4s$, $4p$, and $4d$ electrons as valence electrons. 
For Se, we include $4s$, $4p$, and $4d$ electrons as valence electrons.
For self-consistent calculations, we use an $8{\times}8{\times}6$ 
$k$-point sampling in the three-dimensional (3D) Brillouin zone (BZ) 
for bulk PdSe$_2$ and an $8{\times}8{\times}1$ $k$-point sampling in the
2D BZ for monolayer and bilayer PdSe$_2$. 
The density of states (DOS) is calculated with a finer $k$ grid of 
$20{\times}20{\times}15$ for bulk and $40{\times}40{\times}1$ for 2D 
layers.

\subsection{{\em GW} calculations}

We performed {\em GW} calculations for QP band structures in monolayer, 
bilayer, and bulk PdSe$_2$ using the BERKELEYGW 
code \cite{Deslippe2012,Hybertsen1986,Rohlfing2000}.
We used the one-shot {\em GW} method ({\em G}$_0${\em W}$_0$) which constructs the 
dielectric matrix and the Green's function from DFT eigenvalues
and wave functions and calculates the self-energy once. 
The generalized plasmon-pole model is used to consider the frequency 
dependence of the inverse dielectric function $\epsilon^{-1}(\omega)$.

In {\em GW} calculations, summation over unoccupied bands is needed to obtain
the self-energy. 
Generally, it is known that a large number of unoccupied bands and 
a large energy cutoff for the dielectric matrix are required in {\em GW} 
calculations involving semicore $d$ states, 
while simple $sp$ semiconductors show fast convergence \cite{Shih2010}. 
For calculations of PdSe$_2$, Pd has fully occupied $4d$ states, 
so its semicore $4s$ and $4p$ states should be considered as valence electrons in pseudopotentials.
We include 3900 unoccupied bands in our {\em GW} calculations for bulk PdSe$_2$,
which is about 65 times the number of occupied bands,
and 1050 (2100) unoccupied bands for monolayer (bilayer)
PdSe$_2$, which is about 35 times the number of occupied bands.
For better convergence, we used the static remainder method that 
considers the contribution of unincluded unoccupied bands to the expectation value 
of the static Coulomb-hole operator \cite{Deslippe2013}.
Kinetic-energy cutoffs of 40 and 20~Ry are used to calculate dielectric 
matrices of the bulk and layer, respectively. 
A $4{\times}4{\times}3$ $q$-point sampling is used for the bulk, and for monolayer 
and bilayer PdSe$_2$, a $6{\times}6{\times}1$ uniform $q$ grid is used along 
with additional 10 $q$ points determined by the nonuniform neck subsampling
method \cite{Jornada2017}, which is equivalent to 
a $1143{\times}1143{\times}1$ uniform $q$ grid. 
With these calculation parameters, QP band gaps are converged within a few tens 
of meV.

In general, VBMs and CBMs may not be at a high-symmetry point in BZ.
Thus, to find the QP band gap, which can be indirect, we interpolate 
the QP band structure using the dual-grid method \cite{Rohlfing2000}.
In the dual-grid method, the expectation value of an observable for a wave function in a fine $k$ grid is 
calculated by summing its expectation values for wave functions in a coarse $k$ grid
weighted by coefficients expanding the fine-$k$-grid wave function with 
coarse-$k$-grid wave functions.
QP DOS is also calculated by interpolating 
QP energies from a coarse $k$ grid to a fine $k$ grid.
In our present work, the coarse $k$ grid is $4{\times}4{\times}3$ for bulk and $6{\times}6{\times}1$ for 2D layers, and the fine $k$ grid is
$20{\times}20{\times}15$ for bulk and $40{\times}40{\times}1$ for 2D layers

\section{\label{sec3pdse2}RESULTS AND DISCUSSIONS}

We performed DFT and {\em GW} calculations of monolayer, bilayer, 
and bulk PdSe$_2$ to find direct and indirect band gaps, effective masses, work functions,
ionization potentials, and electron affinities. 
We also obtained the projected density of states to analyze 
contributions of Pd and Se orbitals to electronic states near VBMs and CBMs. 
Finally, we obtained equienergy lines at 0.1~eV 
below the VBM and 0.1~eV above the CBM to find the shape of hole and electron
pockets in the $k$-space.

\subsection{DFT band structures}

\begin{figure}
\includegraphics[width=0.48\textwidth]{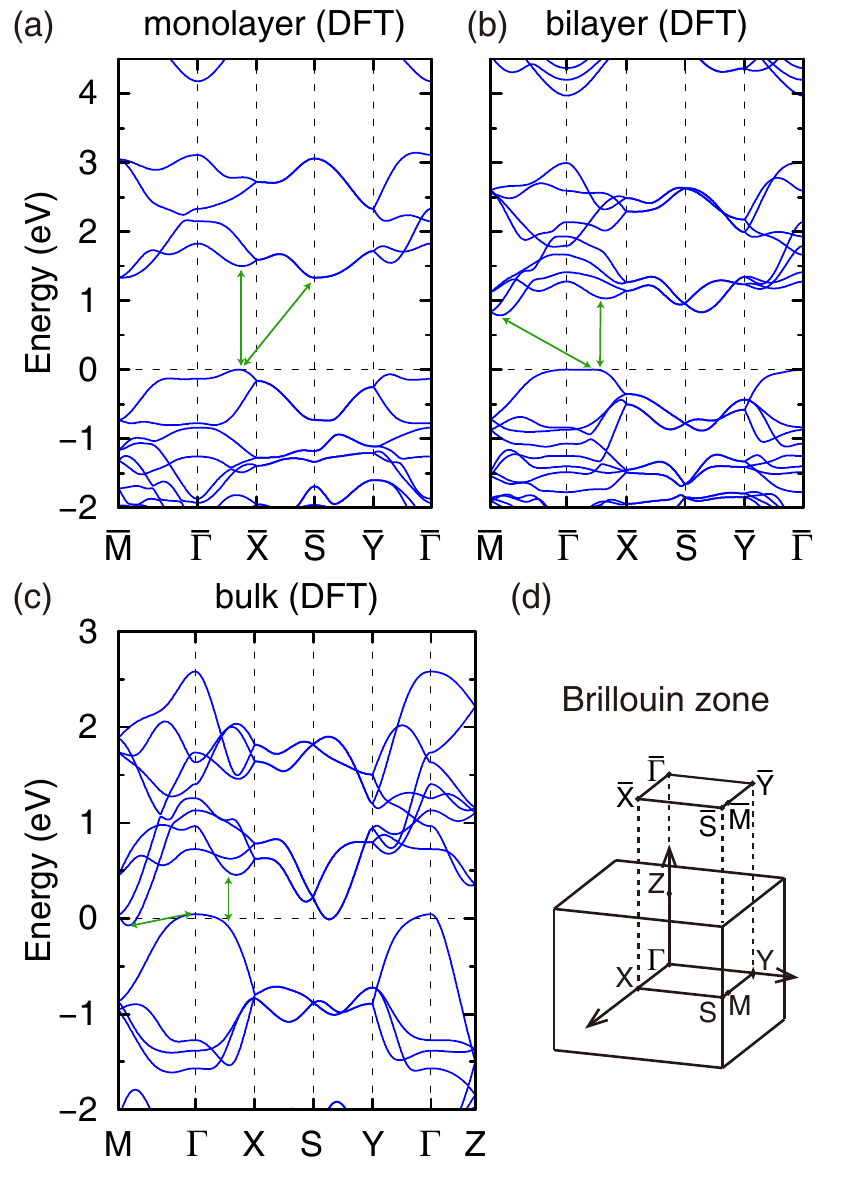}
\caption{\label{fig2pdse2} Band structures of (a) monolayer, (b) bilayer, and
(c) bulk PdSe$_2$ from DFT calculations, and (d) 3D BZ of bulk PdSe$_2$ and
projected 2D BZ of monolayer and bilayer PdSe$_2$.
The $M$ point is not a high-symmetry point, but it is $(t, 0.50, 0.00)$, where $t$ is $0.46$, $0.40$, and $0.42$ for monolayer, bilayer, and bulk PdSe$_2$, respectively.
In (a)-(c), green arrows connect the VBM and CBM for direct and indirect band gaps.}
\end{figure}

Figure~\ref{fig2pdse2} shows electronic band structures for monolayer, bilayer, and 
bulk PdSe$_2$ obtained by DFT calculations.  
Some VBMs and CBMs are located away from high-symmetry lines in the BZ.
Band gaps in monolayer, bilayer, and bulk PdSe$_2$ are all indirect, 
unlike other TMDs in which the band gap is direct or indirect,
depending on the number of layers \cite{Mak2010, Zhang2014}.

Monolayer PdSe$_2$ has an indirect band gap of 1.33~eV, with
the VBM at $(\pm0.34,0,0)$ and the CBM at $(\pm0.44,\pm0.48,0)$ [Fig.~\ref{fig2pdse2}(a)].
Here, $k$-point coordinates are with respect to reciprocal lattice vectors.
The work function is 4.71~eV, which we estimate as the energy 
between the vacuum level and the center of the indirect band 
gap because the chemical potential of an undoped 
semiconductor at low temperature is equal to the average of CBM 
and VBM energies \cite{Kim2021}. 
The ionization potential is 5.37~eV, and the electron affinity is 4.04~eV.
Near the VBM, the effective mass is $0.8m_e$ along the $x$ direction
and $1.3m_e$ along the $y$ direction. Here, $m_e$ is the bare mass
of an electron in vacuum.
Near the CBM, the effective mass is $4.6m_e$ along the $x$ direction
and $0.5m_e$ along the $y$ direction.
The direct band gap is 1.50~eV, located at $(\pm0.36,0,0)$ [Fig.~\ref{fig2pdse2}(a)].
Our DFT indirect band gap 
of monolayer PdSe$_2$ agrees well with results of
previous DFT calculations \cite{Lebegue2013, Sun2015}.

Bilayer PdSe$_2$ has the VBM at $(\pm0.22,0,0)$ and the CBM at $(\pm0.36,\pm0.43,0)$,
exhibiting an indirect band gap of 0.79~eV [Fig.~\ref{fig2pdse2}(b)].
Near the VBM, the valence band shows almost flat band dispersion along the $x$ direction.
The VBM at $(\pm0.22,0,0)$ is 5 $\mu$eV higher than the energy of the highest valence band at the $\Gamma$ point.
In the line from $(\pm0.22,0,0)$ to the $\Gamma$ point, the highest valence band 
has a local minimum, and it is about 4 meV lower than the VBM.
Near the VBM, the effective mass is $2.2m_e$ along the $x$ direction
and $1.4m_e$ along the $y$ direction.
Near the CBM, the effective mass is $0.6m_e$ along the $x$ direction
and $0.3m_e$ along the $y$ direction.
The work function is 4.65~eV, the ionization potential is 5.06~eV, and the electron affinity is 4.27~eV.
The direct band gap is 1.05~eV at $(\pm0.28,0,0)$ [Fig.~\ref{fig2pdse2}(b)].

Bulk PdSe$_2$ is semimetallic with a negative indirect band gap 
of $-$0.10~eV [Fig.~\ref{fig2pdse2}(c)]; that is, the VBM 
is higher than the CBM by 0.10~eV. 
The VBM is at the $\Gamma$ point, and the CBM is at $(\pm0.37,\pm0.44,0)$.
The effective mass near the VBM is $4.1m_e$ along the $x$ direction
and $2.0m_e$ along the $y$ direction.
The effective mass near the CBM is $0.3m_e$ along the $x$ direction
and $0.3m_e$ along the $y$ direction.
The direct band gap is 0.56~eV at $(\pm0.25,0,0)$ [Fig.~\ref{fig2pdse2}(c)].
We obtained band energies of bulk PdSe$_2$ 
with respect to the vacuum level by extracting the vacuum level from 
DFT calculation of 12-layer PdSe$_2$, which is thick enough to have 
a band gap very close to that of bulk. This method was used in the study of
bulk TMDs \cite{Kim2021}.
From band energies with respect to the vacuum level,
the work function of bulk PdSe$_2$ is estimated to be 4.55~eV.

From the resistivity measurement, bulk PdSe$_2$ is a semiconductor 
with a band gap of 0.4~eV \cite{Hulliger1965}, 
while our DFT calculation yields a negative indirect band gap of $-0.1$~eV. 
This underestimation of the band gap in PdSe$_2$ in DFT calculations
may originate from overestimation of the interlayer interaction.
Previous theoretical calculations reported that, without van der Waals 
forces, the relaxed out-of-plane lattice constant $c$ is 
10\% larger than the experimental 
value \cite{Oyedele2017}, and with this lattice constant, 
the DFT calculation gives a band gap of 0.4~eV \cite{Oyedele2017}, 
which is close to the experiment although the atomic structure is not. 
A similar tendency has also been reported for black phosphorus \cite{Baik2015}.

\subsection{QP band structures from {\em GW} calculations}

We obtained QP band structures of PdSe$_2$ from {\em GW} calculations. 
Since the CBM is not in a high-symmetry line in the BZ,
it is necessary to calculate band-edge energies through the dual-grid interpolation 
in order to determine the indirect band gap. 
We found the VBM and CBM of QP band structures by interpolating the QP correction near locations of conduction and valence band edges in DFT band structures.
Then we determined the chemical potential using interpolated QP band structures.
In valence and conduction bands of all the considered cases, 
obtained {\em GW} corrections are fairly smooth with energy, and QP bandwidths 
are slightly larger than DFT bandwidths.
Thus, effective masses in QP band structures are slightly smaller than those in DFT band structures, as shown in Table~\ref{tab1pdse2}.
{\em GW} corrections are almost constant shifts near conduction and 
valence band edges, and band gaps 
are corrected mainly by {\em GW} corrections for valence bands.

\begin{table}
\caption{\label{tab1pdse2} Indirect band gaps $E^{(i)}_g$,
direct band gaps $E^{(d)}_g$, effective masses, ionization potentials (IPs), electron affinities (EAs), and work functions $W$
of monolayer (ML), bilayer (BL), and bulk PdSe$_2$ from DFT and {\em GW} calculations.
Effective masses $m^*_{e,x}$, $m^*_{e,y}$, $m^*_{h,x}$, and $m^*_{h,y}$ 
are for electrons and holes along the $x$ and $y$ axes, respectively.
Band gaps, IP, EA, and $W$ are in eV. 
Effective masses are in units of the bare electron mass in vacuum.
}
\renewcommand{\arraystretch}{1.3}
\begin{tabular}{c c c c c c c c c c c}
\hline
\hline
{ }&{ }&$E^{(i)}_g$&$E^{(d)}_g$&
$m^*_{e,x}$ &$m^*_{e,y}$ &$m^*_{h,x}$ &$m^*_{h,y}$ &IP&EA&$W$ \\
\hline
\multirow{2}*{ML}
& DFT & 1.33    & 1.50    & 4.6 & 0.5 & 0.8 & 1.3 &5.37 &4.04&4.71   \\
& {\em GW} & 2.37   & 2.54   & 4.3 & 0.5 & 0.8 & 1.3 &6.44 &4.06&5.26   \\
\hline
\multirow{2}*{BL}&DFT&0.79   & 1.05   & 0.6 & 0.3 & 2.2 & 1.4 &5.06&4.27& 4.65   \\
& {\em GW} & 1.55   & 1.81   & 0.6 & 0.3 & 2.1 & 1.4 &6.04&4.48& 5.26   \\
\hline
\multirow{2}*{bulk} & DFT & $-$0.10   & 0.56   & 0.3 & 0.3 & 4.1 & 2.0 &&& 4.55 \\
& {\em GW} & 0.45   & 1.10   & 0.3 & 0.3  & 3.9 & 1.9 &5.41&4.96& 5.18\\
\hline
\hline
\end{tabular}
\end{table}

\begin{figure}
\includegraphics[width=0.48\textwidth]{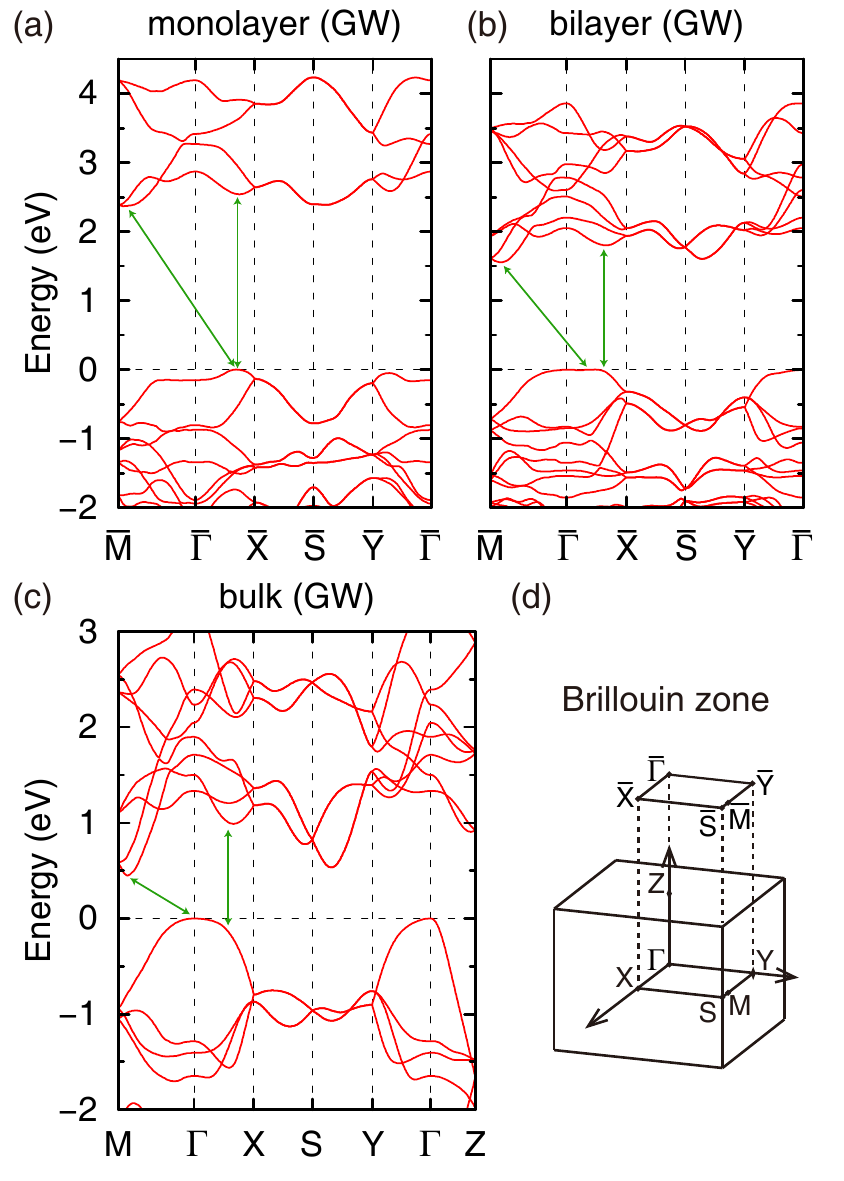}
\caption{\label{fig3pdse2}QP band structures of (a) monolayer, (b) bilayer, 
and (c) bulk PdSe$_2$ from {\em GW} calculations and (d) 3D BZ of bulk PdSe$_2$ and 
projected 2D BZ of monolayer and bilayer PdSe$_2$. 
The $M$ point is not a high-symmetry point, but it is $(t, 0.50, 0.00)$, where $t$ is $0.42$, $0.40$, and $0.40$ for monolayer, bilayer, and bulk PdSe$_2$, respectively. 
In (a)-(c), green arrows connect the VBM and CBM for direct and indirect band gaps.}
\end{figure}

Figure~\ref{fig3pdse2}(a) shows the QP band structure of monolayer PdSe$_2$ obtained from
{\em GW} calculations.
Monolayer PdSe$_2$ has an indirect QP band gap of 2.37~eV,
with the VBM at $(\pm0.34,0,0)$ and the CBM at $(\pm0.39,\pm0.47,0)$.
{\em GW} correction is much larger in monolayer PdSe$_2$ than in the bulk.
For monolayer PdSe$_2$, a large band-gap 
correction leads to a large work-function change. 
While the work function from DFT calculation is 4.71~eV for 
the monolayer, {\em GW} calculations increase this value to 5.26 eV 
for the monolayer.
The ionization potential is 6.44~eV, and the electron affinity is 4.06~eV. 
The {\em GW} correction for band-edge energies is prominent in valence bands.
Near the VBM, the effective mass is $0.8m_e$ along the $x$ direction
and $1.3m_e$ along the $y$ direction. 
Near the CBM, the effective mass is $4.3m_e$ along the $x$ direction
and $0.5m_e$ along the $y$ direction. The effective masses are slightly reduced by {\em GW} corrections when compared with DFT results.

Figure~\ref{fig3pdse2}(b) shows the QP band structure of bilayer PdSe$_2$ obtained from 
{\em GW} calculations.
Bilayer PdSe$_2$ has an indirect QP band gap of 1.55~eV,
with the VBM at $(\pm0.24,0,0)$ and the CBM at $(\pm0.34,\pm0.43,0)$.
{\em GW} correction is much larger in bilayer PdSe$_2$ than in the bulk.
For bilayer PdSe$_2$, a large band-gap 
correction leads to a large work-function change. 
While the work function from the DFT calculation is 4.65~eV for the bilayer, 
the {\em GW} calculation increases this value to 5.26 eV for the bilayer, 
which is the same value as that for the monolayer.
The ionization potential is 6.04~eV, and the electron affinity is 4.48~eV. 
The {\em GW} correction for band-edge energies is prominent in the valence bands as in the monolayer.
Near the VBM, the effective mass is $2.1m_e$ along the $x$ direction
and $1.4m_e$ along the $y$ direction. 
Near the CBM, the effective mass is $0.6m_e$ along the $x$ direction
and $0.3m_e$ along the $y$ direction. The effective masses are slightly reduced by {\em GW} corrections when compared with DFT results as in the monolayer.

Figure~\ref{fig3pdse2}(c) shows the QP band structure of bulk PdSe$_2$ obtained from 
{\em GW} calculations.
Bulk PdSe$_2$ has an indirect QP band gap of 0.45~eV,
with the VBM at $\Gamma$ and the CBM at $(\pm0.35,\pm0.44,0)$.
The calculated QP band gap of 0.45~eV is in excellent agreement with the experimental band gap of 0.4~eV from the resistivity measurement \cite{Hulliger1965}.
The {\em GW} correction in bulk PdSe$_2$ is about $-0.4$ and $-0.9$~eV 
for conduction and valence bands, respectively, increasing the band gap 
by about 0.5 eV and increasing the work function.
The {\em GW} correction in bulk PdSe$_2$ is much smaller than in the monolayer and bilayer,
but the correction is large enough to change the semimetallic DFT band
structure to the semiconducting one.
The work function is 5.18~eV, the ionization potential is 5.41~eV, and the electron affinity is 4.96~eV. 
Near the VBM, the effective mass is $3.9m_e$ along the $x$ direction
and $1.9m_e$ along the $y$ direction. 
Near the CBM, the effective mass is $0.3m_e$ along the $x$ direction
and $0.3m_e$ along the $y$ direction. The effective masses are slightly reduced by {\em GW} corrections when compared with DFT results as in the monolayer and bilayer.

\begin{figure}
\includegraphics[width=0.48\textwidth]{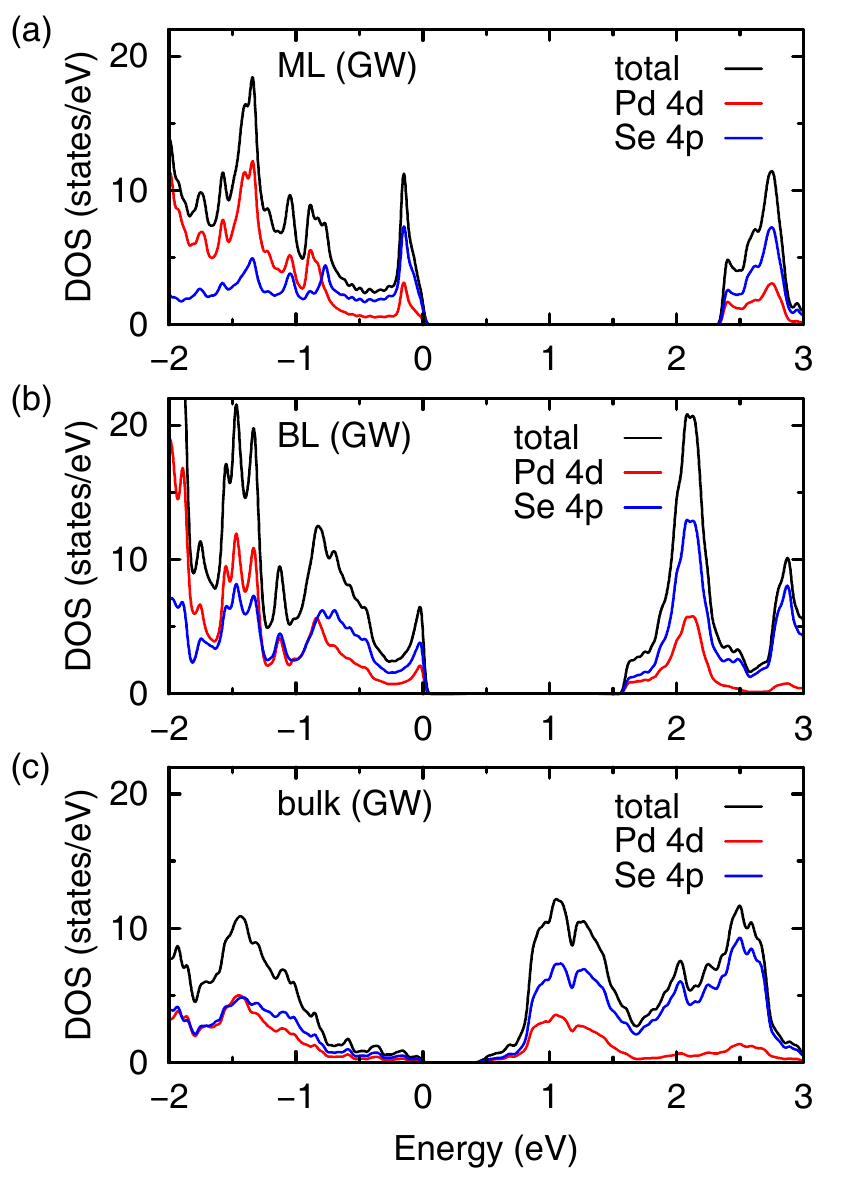}
\caption{\label{fig4pdse2}QP DOS of (a) monolayer, (b) bilayer, 
and (c) bulk PdSe$_2$ from {\em GW} calculations. 
QP energies are interpolated to a dense $k$ grid to calculate DOS.
States near the CBM and VBM are mainly from Pd $4d$ and Se $4p$ orbitals. }
\end{figure}

\subsection{Density of states}

Figure~\ref{fig4pdse2} shows the QP DOS for monolayer, bilayer, and bulk PdSe$_2$ 
obtained from {\em GW} calculations.
We obtained QP band energies for a coarse $k$-point grid by {\em GW} calculations 
and then interpolated them to a fine $k$-point grid to calculate the QP DOS.
In Figs.~\ref{fig4pdse2}(a)-\ref{fig4pdse2}(c), the VBM is set to zero in energy. 
Projected QP DOSs show that states near the VBM and CBM 
are mainly from Pd $4d$ and Se $4p$ orbitals, and contributions of Se $4p$ orbitals to states near the VBM and CBM are greater
than those of Pd $4d$ orbitals. This implies that 
each Se or Pd atom contributes similarly to the QP DOS near the VBM and CBM, 
for Se atoms are twice as numerous as Pd atoms in the material.
Thus, hybridization of Pd $4d$ orbitals and Se $4p$ orbitals is important 
for describing the band gap in PdSe$_2$. 

In Fig.~\ref{fig4pdse2}(a), the QP DOS of monolayer PdSe$_2$ shows a very sharp peak 
right below the VBM originating from almost flat bands near the VBM shown in Fig.~\ref{fig3pdse2}(a).
In Fig.~\ref{fig4pdse2}(b), the QP DOS of bilayer PdSe$_2$ shows a peak near the VBM which is not as
strong as that in the monolayer but closer to the VBM in energy.
In contrast, the QP DOS of bulk PdSe$_2$, shown in Fig.~\ref{fig4pdse2}(c), does not have a peak
near the VBM. Thus, with hole doping, ferromagnetic instability may be less 
effective in the bulk than in the monolayer and bilayer.

\begin{figure}
\includegraphics[width=0.48\textwidth]{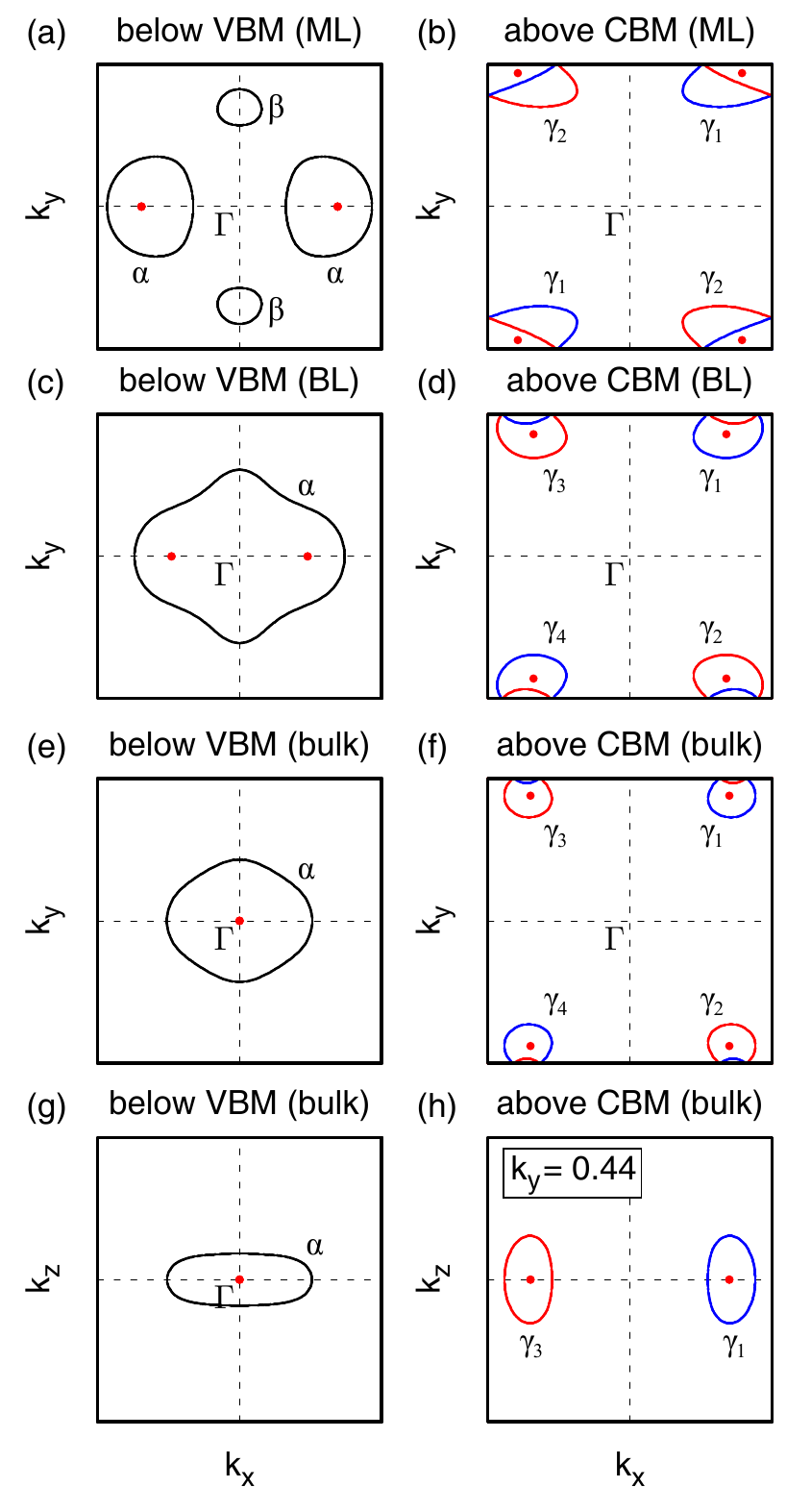}
\caption{\label{fig5pdse2}Equienergy lines at 0.1~eV below the VBM in (a) monolayer, 
(c) bilayer, and (e) and (g) bulk, and at 0.1~eV above the CBM
in (b) monolayer, (d) bilayer, and (f) and (h) bulk
obtained with {\em GW} calculations.
In (a)-(f), equienergy lines are plotted in the $k_xk_y$ plane.
In (g), equienergy lines are plotted in the $k_y = 0$ plane.
In (h), equienergy lines are plotted in the $k_y = 0.44$ plane.
In each plot, red dots represent locations of the VBM or CBM.}
\end{figure}

\subsection{Equienergy lines in the {\em k} space}

Figure~\ref{fig5pdse2} shows equienergy lines in the BZ obtained from {\em GW} calculations of monolayer, 
bilayer, and bulk PdSe$_2$ at energies which are 0.1~eV below 
the VBM and 0.1~eV above the CBM.
These lines correspond to hole or electron pockets in the BZ if hole or electron 
doping is done.

Monolayer PdSe$_2$ has the VBM at $(\pm0.34,0,0)$ and 
the CBM at $(\pm0.39,\pm0.47,0)$, as marked with red dots
in Figs.~\ref{fig5pdse2}(a) and \ref{fig5pdse2}(b), respectively.
In monolayer PdSe$_2$, equienergy lines 0.1~eV below the VBM form four close loops,
two of which (marked with $\alpha$) are around the VBM and larger than the other 
two (marked with $\beta$) [Fig.~\ref{fig5pdse2}(a)].
Meanwhile, equienergy lines 0.1~eV above the CBM form loops near four corners of the BZ [Fig.~\ref{fig5pdse2}(b)],
which are two intersecting slanted ellipses (marked with $\gamma_1$ and $\gamma_2$)
when all BZ corners are put together.

Bilayer PdSe$_2$ has the VBM at $(\pm0.24,0,0)$ and the CBM 
at $(\pm0.34,\pm0.43,0)$, as marked with red dots
in Figs.~\ref{fig5pdse2}(c) and \ref{fig5pdse2}(d), respectively.
In bilayer PdSe$_2$, the equienergy line 0.1~eV below the VBM is a large 
loop (marked with $\alpha$)
surrounding two VBM points [Fig.~\ref{fig5pdse2}(c)], indicating that its valence band 
is quite flat near the VBM.
Meanwhile, equienergy lines 0.1~eV above the CBM are loops near four corners of the BZ,
which form four closed loops (marked with $\gamma_1$, $\gamma_2$, $\gamma_3$,
and $\gamma_4$) when horizontal BZ boundaries are put together 
[Fig.~\ref{fig5pdse2}(d)]. Compared with the monolayer, VBM points are shifted a little bit 
toward the $\Gamma$ point, and CBM points are shifted a little bit away from
BZ corners.

Bulk PdSe$_2$ has the VBM at $\Gamma$ and the CBM at $(\pm0.35,\pm0.44,0)$, 
as marked with red dots in Figs.~\ref{fig5pdse2}(e) and \ref{fig5pdse2}(f), respectively.
In bulk PdSe$_2$, the equienergy line 0.1~eV below the VBM is a closed loop (marked
with $\alpha$) surrounding the $\Gamma$ point in the $k_xk_y$ plane [Fig.~\ref{fig5pdse2}(e)], 
indicating that the in-plane effective mass is large.
Meanwhile, equienergy lines 0.1~eV above the CBM are loops near four corners of the
BZ, which form four closed loops (marked with $\gamma_1$, $\gamma_2$, $\gamma_3$,
and $\gamma_4$) when horizontal BZ boundaries are put together
[Fig.~\ref{fig5pdse2}(f)], similar to the case of the bilayer.

Since the BZ of the bulk is three-dimensional, we also plot the equienergy line
0.1~eV below the VBM in the $k_xk_z$ plane, that is, the $k_y = 0$ 
plane [Fig.~\ref{fig5pdse2}(g)], and equienergy lines 0.1~eV above the CBM in the $k_y = 0.44$ 
plane [Fig.~\ref{fig5pdse2}(h)]. In the $k_y = 0$ plane, the equienergy line 0.1~eV 
below the VBM is a very narrow ellipselike loop corresponding 
to $\alpha$ [Fig.~\ref{fig5pdse2}(g)], indicating that the effective mass is much 
smaller along the $z$ direction than the $x$ direction.
In the $k_y = 0.44$ plane which passes through two CBM points, equienergy 
lines 0.1~eV above the CBM are two ellipselike loops corresponding 
to $\gamma_1$ and $\gamma_3$ [Fig.~\ref{fig5pdse2}(h)], indicating that the 
effective mass is larger along the $z$ direction than the $x$ direction.

\section{\label{sec4pdse2}SUMMARY}

We studied the electronic structures of 
monolayer, bilayer, and bulk PdSe$_2$ using first-principles 
DFT and {\em GW} calculations. 
First, we calculated the electronic structure of bulk PdSe$_2$ using DFT, 
obtaining a semimetallic electronic structure due to underestimation 
of the band gap. To calculate the band gap of PdSe$_2$ correctly, 
we performed {\em GW} calculations. We obtained the QP band structure of bulk 
PdSe$_2$ with an indirect band gap of 0.45~eV, which is consistent with 
the reported experimental band gap of 0.4~eV.
Then we calculated the electronic structure of 2D PdSe$_2$ with respect to the vacuum level using DFT and the {\em GW} method.
Our DFT calculations produced band gaps of 1.33 and 0.79~eV for monolayer 
and bilayer PdSe$_2$, respectively. From {\em GW} calculations, 
QP band gaps of monolayer and bilayer PdSe$_2$ are 2.37 and 1.55~eV, 
respectively.
We also obtained the DOS, effective masses, work functions, ionization potentials, electron affinities, and $k$-space shapes of electron and hole pockets for monolayer, bilayer, and bulk PdSe$_2$ from DFT and {\em GW} calculations. These results provide basic information for the development of material properties and device applications.
\\

\begin{center}
{\bf ACKNOWLEDGMENTS}\\
\end{center}

This work is supported by NRF of Korea (Grant No. 2020R1A2C3013673). 
Computational resources were provided by the KISTI Supercomputing 
Center (Project No. KSC-2020-CRE-0335).\\

\appendix

\begin{center}
{\bf APPENDIX: DOS FROM DFT CALCULATIONS}
\\
\end{center}

\begin{figure}[b]
\includegraphics[width=0.48\textwidth]{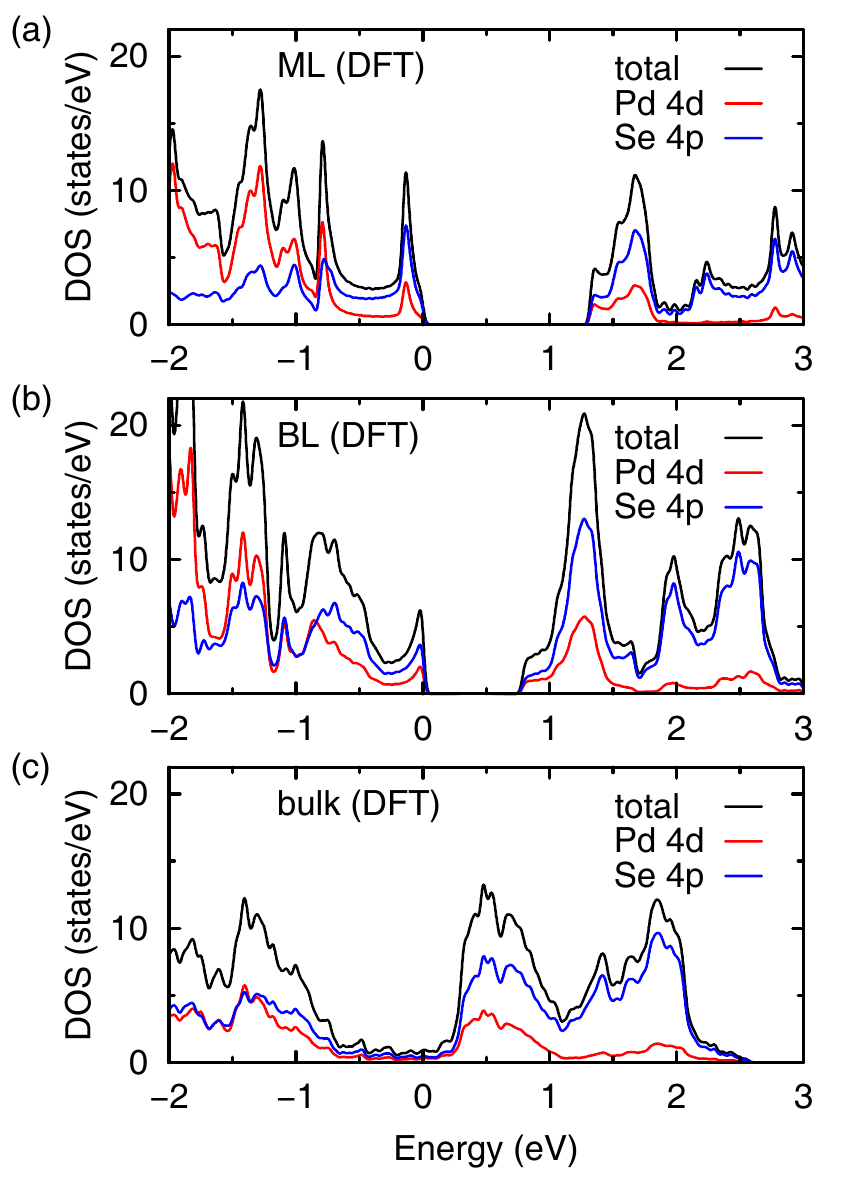}
\caption{\label{fig6pdse2}DOS of (a) monolayer, (b) bilayer, 
and (c) bulk PdSe$_2$ from DFT calculations. 
States near the CBM and VBM are mainly from Pd $4d$ and Se $4p$ orbitals. }
\end{figure}

Figure~\ref{fig6pdse2} shows the electronic DOS of monolayer, bilayer, and bulk PdSe$_2$ obtained from DFT calculations.
In Figs.~\ref{fig6pdse2}(a) and \ref{fig6pdse2}(b), VBM is set to zero in energy.
In Fig.~\ref{fig6pdse2}(c), the chemical potential is set to zero.
Overall shapes of the electronic DOS are not much different from those 
of the QP DOS from {\em GW} calculations, except for sizes of band gaps.
Compared with the QP DOS from {\em GW} calculations,
the DOSs from DFT calculations show smaller band gaps in the monolayer and 
bilayer [Figs.~\ref{fig6pdse2}(a) and \ref{fig6pdse2}(b)] and a negative indirect band gap in the bulk [Fig.~\ref{fig6pdse2}(c)].
Because of the negative indirect band gap in DFT calculations,
DOS in Fig.~\ref{fig6pdse2}(c) does not show any band gap explicitly,
although bulk PdSe$_2$ has a finite energy difference between valence and 
conduction bands throughout the BZ, as shown in Fig.~\ref{fig2pdse2}(c).
Meanwhile, similar to the QP DOS from {\em GW} calculations, the DOSs from DFT calculations 
show sharp peaks near the VBM in the monolayer and bilayer, 
as shown in Figs.~\ref{fig6pdse2}(a) and \ref{fig6pdse2}(b), but no peak near the VBM in the bulk [Fig.~\ref{fig6pdse2}(c)].


\end{document}